\documentclass{jfm}
\usepackage{graphicx}
\usepackage{epstopdf, epsfig}
\usepackage{tikz}
\usetikzlibrary{positioning}

\usepackage[normalem]{ulem}

\shorttitle{Regime crossover in Rayleigh-B\'enard convection}
\shortauthor{Rodolfo Ostilla-M\'onico, Amit Amritkar}

\title{Regime crossover in Rayleigh-B\'enard convection with mixed boundary conditions}

\author{Rodolfo Ostilla-M\'onico\aff{1}
  \corresp{\email{rostilla@central.uh.edu}},
  Amit Amritkar\aff{1,2}}

\affiliation{\aff{1}Department of Mechanical Engineering, University of Houston,
Houston, TX 77204, USA
\aff{2}HPE Data Science Institute, University of Houston, 
Houston, TX 77204, USA}

\begin{document}

\maketitle

\begin{abstract}
We numerically simulate three-dimensional Rayleigh-B\'enard convection, the flow in a fluid layer heated from below and cooled from above, with inhomogeneous temperature boundary conditions to explore two distinct regimes described in recent literature. We fix the non-dimensional temperature difference, i.e.~the Rayleigh number to $Ra=10^8$, and vary the Prandtl number between $1$ and $100$. By introducing stripes of adiabatic boundary conditions on the top plate, and making the surface of the top-plate only $50\%$ conducting, we modify the heat transfer, average temperature profiles and the underlying flow properties. We find two regimes: when the pattern wavelength is small, the flow is barely affected by the stripes. The heat transfer is reduced, but remains a large fraction of the unmodified case, and the underlying flow is only slightly modified. When the pattern wavelength is large, the heat transfer saturates to approximately two-thirds of the value of the unmodified problem, the temperature in the bulk increases substantially, and velocity fluctuations in the directions normal to the stripes are enhanced. The transition between both regimes happens at pattern wavelength around the distance between two plates, with different quantities transitioning at slightly different wavelength values. This transition is approximately Prandtl number independent, even if the statistics in the long-wavelength regime slightly vary.
\end{abstract}

\begin{keywords}
%Heat transfer
\end{keywords}

\section{Introduction}

Rayleigh-B\'enard (RB) convection, the flow in a layer heated from below and cooled from above, is a canonical model for the problem of thermal convection \citep{ahl09,loh10,chi12}. The study of Rayleigh-B\'{e}nard convection has proven so fruitful because the system is well-defined, closed and possesses non-trivial conservation properties accessible to theory and experiment, such as the exact relationships between kinetic energy dissipation and heat transport \citep{shr90}. However, most real systems both in process-technology and in Nature differ from the idealized Rayleigh-B\'enard setup. Many modifications of the canonical system are possible, such as the addition of roughness \citep{tis11,rus18}, finite conductivity effects on the plates \citep{ver04,bro05} or different sidewall conditions \citep{xia97,poe14}. A variation which has attracted recent attention is that of inhomogeneous temperature boundary conditions due to its interest in geophysics \citep{pek35, len03, len05, sol00}. To model the difference between continental and oceanic lithospheres, studies have substituted the constant temperature top boundary condition by a pattern of adiabatic and conducting boundary conditions, meant to represent the physical properties of continents and oceans respectively \citep{coo13, rip14, wan17, bak18}. However, the experimental and numerical results for this configuration appear to be at odds with each other \citep{ost17}.

Studies of Rayleigh-B\'enard with mixed temperature boundary conditions were pioneered by the simulations of \cite{coo13}, who studied the effect of large, or small patches of adiabatic ``continents'' in a doubly periodic, large-aspect ratio RB simulation domain (cell). \cite{coo13} found that the heat transfer was largely unaffected by the number of continental blocks, and depended on area coverage. These results were corroborated experimentally by \cite{wan17}, who found that the heat transfer in their rectangular RB cell depended mainly on the conducting area, and was largely independent of the arrangement of the plates. For both studies, the arrangement of the plates had a substantial effect on the underlying flow. The role of plate size and shape was postulated to be crucial in understanding the role continental and oceanic plates play in the configuration of the mantle circulation, and how these interactions determine the dynamics of the Wilson cycle \citep{ost17}.

Conversely, the simulations of \cite{rip14} in two dimensions, and their recent extension by \cite{bak18} to three dimensions, focused on patches which have characteristic dimensions smaller than the distance between plates. These simulations found that the arrangement of the plates, quantified as a stripe wavelength, had a substantial effect on the total heat transport, and that once the stripe wavelength becomes comparable to the size of the thermal boundary layer, the heat transport asymptotically reaches the values corresponding to a fully conducting plate, \emph{even if the conducting area is only one half}. But unlike the studies mentioned above, both of these studies found no significant effect of plate size on the flow dynamics.

The discrepancy between these two results becomes less apparent when one compares the wavelength of the inhomogeneities. The stripes in the first case scenario by \cite{rip14,wan17} are much larger than the stripes in the second case scenario by \cite{rip14,bak18}. This discrepancy just indicates that there are two distinct regimes, and that a cross-over between the two regimes must exist: between large adiabatic patches which significantly affect the flow topology and small adiabatic patches whose effect is only present very close to the plates, there must be a transitional regime with ``medium''-sized patches which affect both types of statistics. Because land masses such as continents and islands come in a wide variety of shapes and length-scales, understanding how this size and shapes affects the circulation of the mantle can enhance our understanding of continental drift.

In this study we set out to find the location and characteristics of the cross-over regime between the ``large'' and ``small'' pattern behaviour, and to explore its characteristics through three-dimensional direct numerical simulations. We note that this transition might not happen at the same stripe size for different statistics, as has been seen for example in the transition between different flow regimes in rotating Rayleigh-B\'enard convection \citep{kun16}. For this, we will simulate three-dimensional Rayleigh-B\'enard with non-uniform temperature boundary conditions on the top surface, which are a mixture of adiabatic and perfectly conducting. 

% fractures, leads and polynyas in ice affect the dynamics of oceanic and atmospheric circulation \citep{may82}, and by extension global warming, is crucial in today’s world where ice caps are disappearing.

\section{Numerical details}

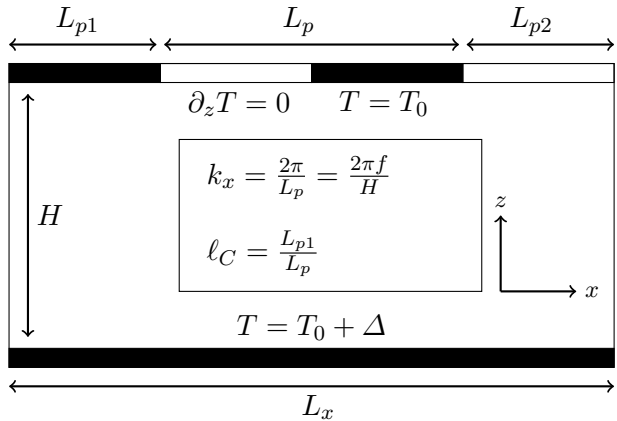
\begin{figure}
    \centering
    \begin{tikzpicture}
    \draw (0,0) rectangle (8,4);
    \draw [fill=black] (0,3.75) rectangle (2,4);
    \draw (2,3.75) rectangle (4,4);
    \draw [fill=black] (4,3.75) rectangle (6,4);
    \draw (6,3.75) rectangle (8,4);
    \draw [fill=black] (0,0) rectangle (8,0.25);
    \node [below right,black] at (3.75,-0.25) {\large $L_x$};
    \draw [<->,thick] (0,-0.25) -- (8,-0.25);
    \node [below right,black] at (0.25,2.25) {\large $H$};
    \draw [<->,thick] (0.25,0.40) -- (0.25,3.60);
    \node [above right,black] at (3.5,4.25) {\large $L_p$};
    \draw [<->,thick] (2.05,4.25) -- (5.95,4.25);
    \node [above right,black] at (0.5,4.25) {\large $L_{p1}$};
    \draw [<->,thick] (0,4.25) -- (1.95,4.25);
    \node [above right,black] at (6.5,4.25) {\large $L_{p2}$};
    \draw [<->,thick] (6.05,4.25) -- (8,4.25);
    \node [below right,black] at (2.25,3.75) {\large $\partial_z T = 0$};
    \node [below right,black] at (4.25,3.75) {\large $T = T_0$};
    \node [above, black] at (4,0.25) {\large $T = T_0+\Delta$};
    \node [right, black] at (2.5,2.5) {\large $k_x = \frac{2 \pi}{L_p} =
    \frac{2 \pi f}{H}$};
    \node [right, black] at (2.5,1.5) {\large $\ell_C = \frac{L_{p1}}{L_{p}}$};
    \draw (2.25,1) rectangle (6.25,3);
    \draw [->,thick] (6.5,1) -- (7.5,1);
    \draw [->,thick] (6.5,1) -- (6.5,2);
    \node [above,black] at (6.5,2) {$z$};
    \node [right,black] at (7.5,1) {$x$};
    \end{tikzpicture}
    \caption{Schematic of the simulation setup. The third dimension is removed for clarity purposes. The bottom plate is set at a constant (hot) temperature, while in the top plate, stripes of adiabatic and conducting boundary conditions alternate every $L_p$. We set $L_{p1}=L_{p2}$ such that the effective area of conducting material is $50\%$. }
    \label{fig:sketch}
\end{figure}

We perform large aspect ratio simulations of RB flow using a second order, centered finite difference code \citep{poe15}. The code is periodic in the wall parallel directions with equal periodicity lengths of $L_x=L_y=L$. We fix the Rayleigh number to $Ra=g\beta\Delta H^3/(\nu \kappa)=10^8$, and vary the Prandtl number $Pr=\nu/\kappa$ between $1$ and $100$, where $g$ is the acceleration of gravity, $\beta$ the fluid thermal expansion coefficient, $\Delta$ the hot-cold temperature difference, $H$ the height of the cell and $\nu$ and $\kappa$ the kinematic  viscosity and thermal diffusivity of the fluid. The temperature boundary conditions are imposed as Dirichlet boundary conditions on the conducting parts of the plates: the bottom plate is fixed to a temperature $\Delta$ above that of the conducting regions of the top plate. In this way, the Rayleigh number $Ra$ can be thought of as a non-dimensional measure of the temperature difference while the Prandtl number is a fluid property.
The 2D adiabatic stripes are introduced in the top plate using Neumann no-flux conditions. This means that the temperature of the fluid close to the adiabatic plates is not determined. See Figure \ref{fig:sketch} for a sketch of the configuration and Figure \ref{fig:tempvis} for a flow visualization.

The non-dimensionalized periodicity of the system $\Gamma=L/H$ is a numerical parameter whose influence we want to remove as much as possible. Previous studies \citep{ste18} have shown that it has a strong influence on the underlying statistics of the flow. We simulate aspect ratios in the range from $\Gamma=L/H=1$ to $\Gamma=16$. As we will discuss below further, an adequately normalized heat transfer, as studied in \cite{bak18} does not show significant box-size/domain-size dependence. The temperature statistics show some box-size dependence, especially when the number of unit patterns in the domain is small. The velocity statistics show a strong box-size dependence, which increases with Prandtl number. For $Pr=1$, the velocity statistics for both horizontal directions are approximately equal $\Gamma\gtrsim 2$. But for $Pr=100$, even at $\Gamma=8$ we could not recover isotropic statistics in both horizontal directions. Therefore, velocity statistics will only be shown for $Pr=1$, where we can be sure that any anisotropy between the horizontal directions is produced by the inhomogeneous boundary conditions instead of by numerical effects.

To fully capture the cross-over regime, we simulate a wide range of stripe wave numbers $k_x=2\pi/L_p$, where $L_p=H/f$ is the stripe wavelength, and $f$ the number of stripes per unit non-dimensional length (non-dimensionalized using the height). We set $L_{p1}=L_{p2}$, and keep the top plate equally partitioned between conducting and adiabatic regions. The largest wavelength covered is $f=1/8$, representing a single stripe pair in a $\Gamma=8$ domain. This wavelength is larger than that of the experiments by \cite{wan17}, and will cover the full experimental parameter regime. The smallest wavelength is $f=90$, well into the asymptotic small-patch region according to \cite{bak18}. Due to the second-order scheme of the code used, each stripe must be resolved by at least four points, or otherwise artifacts are introduced. This means that to keep on increasing $f$, we would have to increase the resolution. Because of this, we limited ourselves to $f\leq 90$. Furthermore, we note that for some values of $f$, we have simulated several periodicity aspect ratios $\Gamma$ to quantify the domain independence. A full list of all the simulated $\Gamma$ and $f$ is available in Table \ref{tbl:gammaf}.

\begin{table}
    \centering
\begin{tabular}{ c | l}
$\Gamma$ & $f$ \\ 
 1 & 1,2,4,6,10,15,20,30,45,60,90 \\
 2 & 0.5,1,2,4,10 \\
 4 & 0.25,0.5,1,2,4 \\
 8 & 0.125,0.25,0.5\\
 16 & 0.125  (Only $Pr=1$)
\end{tabular}
    \caption{Simulated values of $f$ and $\Gamma$}
    \label{tbl:gammaf}
\end{table}

We will focus on one-dimensional stripes on the plates, as our previous exploration of checkerboard patterns showed that they do not produce significantly different physics \citep{bak18}. However, stripes introduce an asymmetry between both horizontal directions which can affect the flow enhancement. We have also simulated two cases with checkerboard inhomogeneities to better quantify the effects of boundary asymmetry against large, but isotropic inhomogeneities.

The grid resolution used for this study conforms to the grid convergence study performed by \cite{bak18} for $Pr=1$. For larger Prandtl numbers, we use the same resolution, and check its adequacy by monitoring the balance between viscous dissipation and heat transport outlined in \cite{ste10}. Finally, temporal convergence is assessed through measuring the non-dimensional heat transport, i.e. the Nusselt number $Nu=Q/(\kappa\Delta H^{-1})$, with $Q$ the plate-to-plate heat transfer. While $Q$ can be measured in many ways \citep{shr90}, we use the volumetrically- and temporally averaged value of $\langle u_z \theta \rangle$ to obtain the numerical value of $Q$. As this is a second-order correlation which takes longer to converge, it provides some bounds on our temporal convergence errors. We have ran the simulations for small domains with $\Gamma=1$ up to $1200$ large-eddy turnover times, and the larger domains at $\Gamma=8$ up to $120$ large-eddy turnover times. This gives us an error bound on temporal convergence of less than $1\%$. 

\section{Results}

\begin{figure}
 \includegraphics[trim={3cm 6cm 3cm 10cm},clip,width=0.48\linewidth]{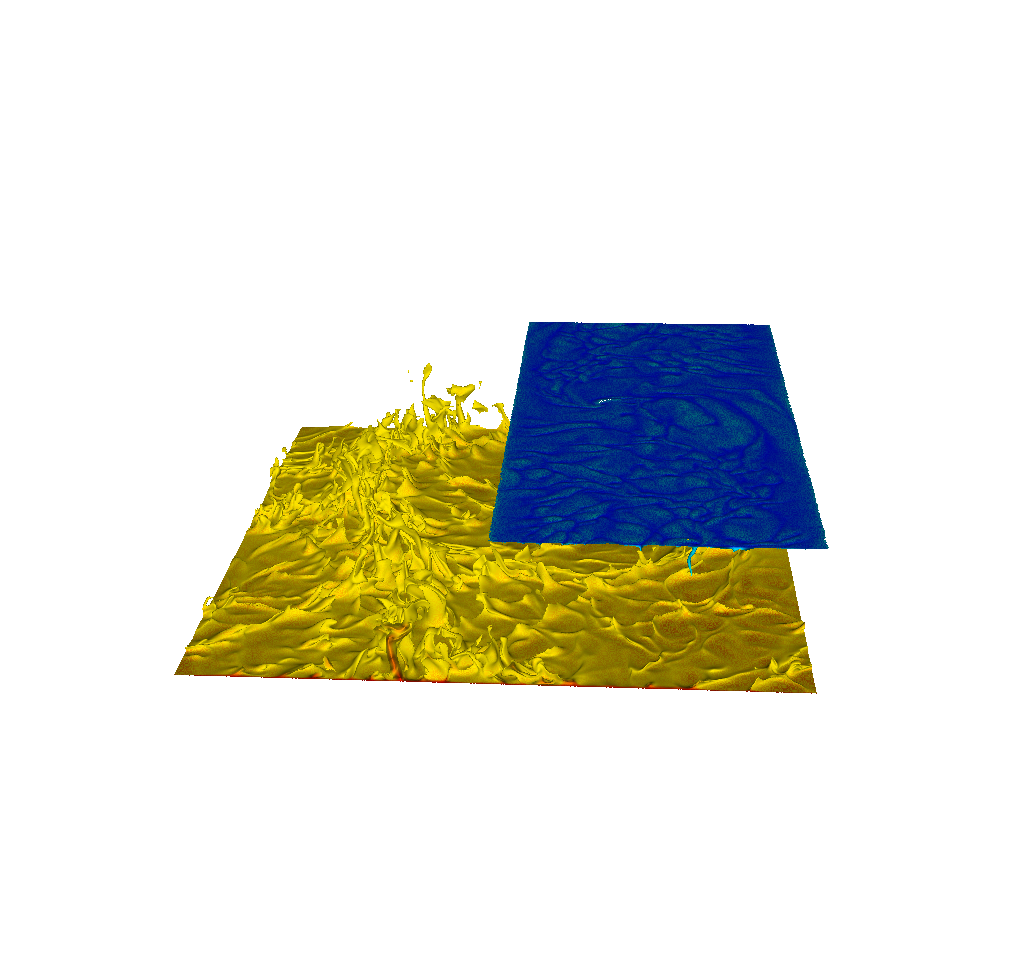}
 \includegraphics[trim={3cm 6cm 3cm 10cm},clip,width=0.48\linewidth]{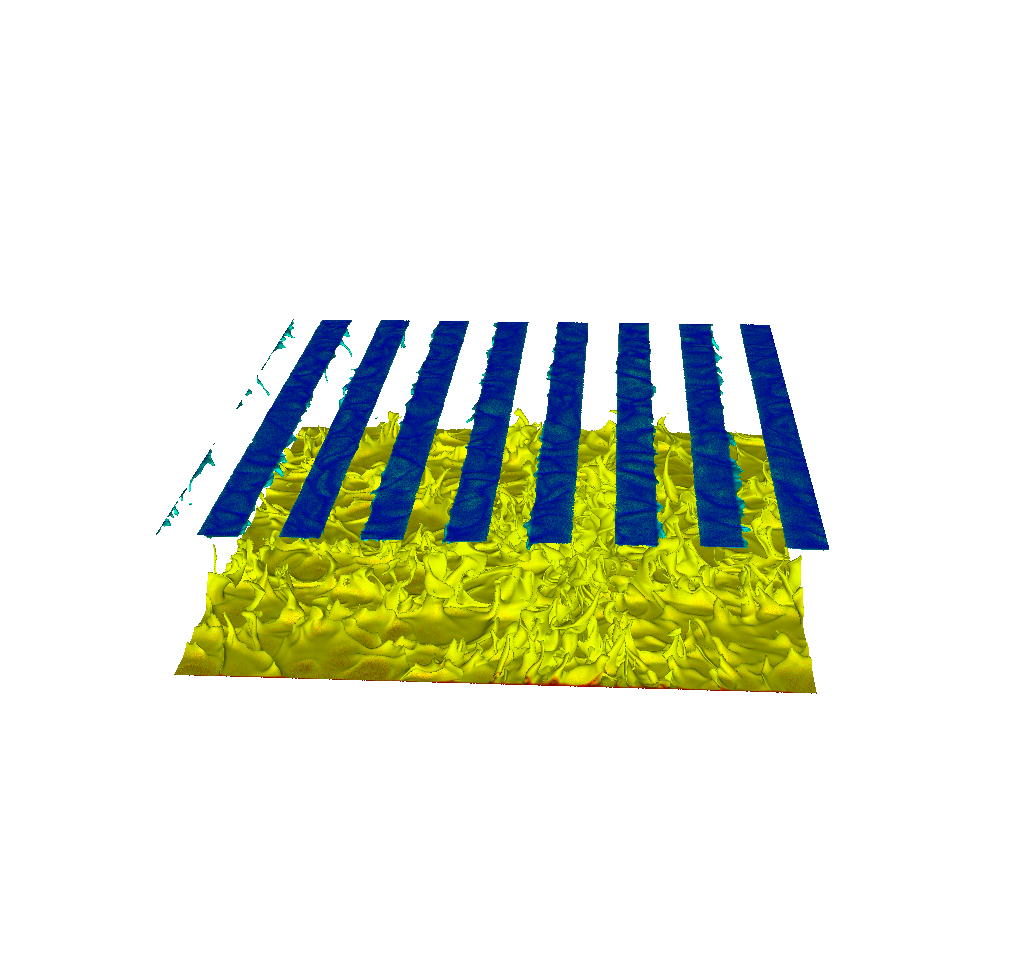}
 \caption{Volume rendering visualization of the instantaneous temperature for two cases simulated at $Ra=10^8$, $Pr=1$ and $\Gamma=4$. Red denotes hot fluid while blue denotes cold fluid. Left: Large stripes with $f=0.25$, the stripes cause a substantial ordering of the plume dynamics in the bottom plate, and an overall increase in temperature. Right: Small stripes with $f=2$. The stripes do not affect the flow dynamics at the bottom plate. }
 \label{fig:tempvis}
\end{figure}

We start by showing a visualization of the instantaneous temperature field for two cases with different values of $f$ in Figure \ref{fig:tempvis}. From a cursory glance, we can confirm what was detailed in the introduction: two very different regimes exist; one where the stripes are small and the flow dynamics is not significantly affected as compared to fully conducting top plate case (shown on the right), and one where the stripes are large and the flow dynamics is substantially changed due to the presence of a large adiabatic surface only on one side (shown on the left). The rest of this manuscript will involve teasing out the differences between both flow regimes, when the transition between them happens, and how they affect the flow statistics.

\subsection{Heat transfer}

\begin{figure}
 \includegraphics[width=0.48\linewidth]{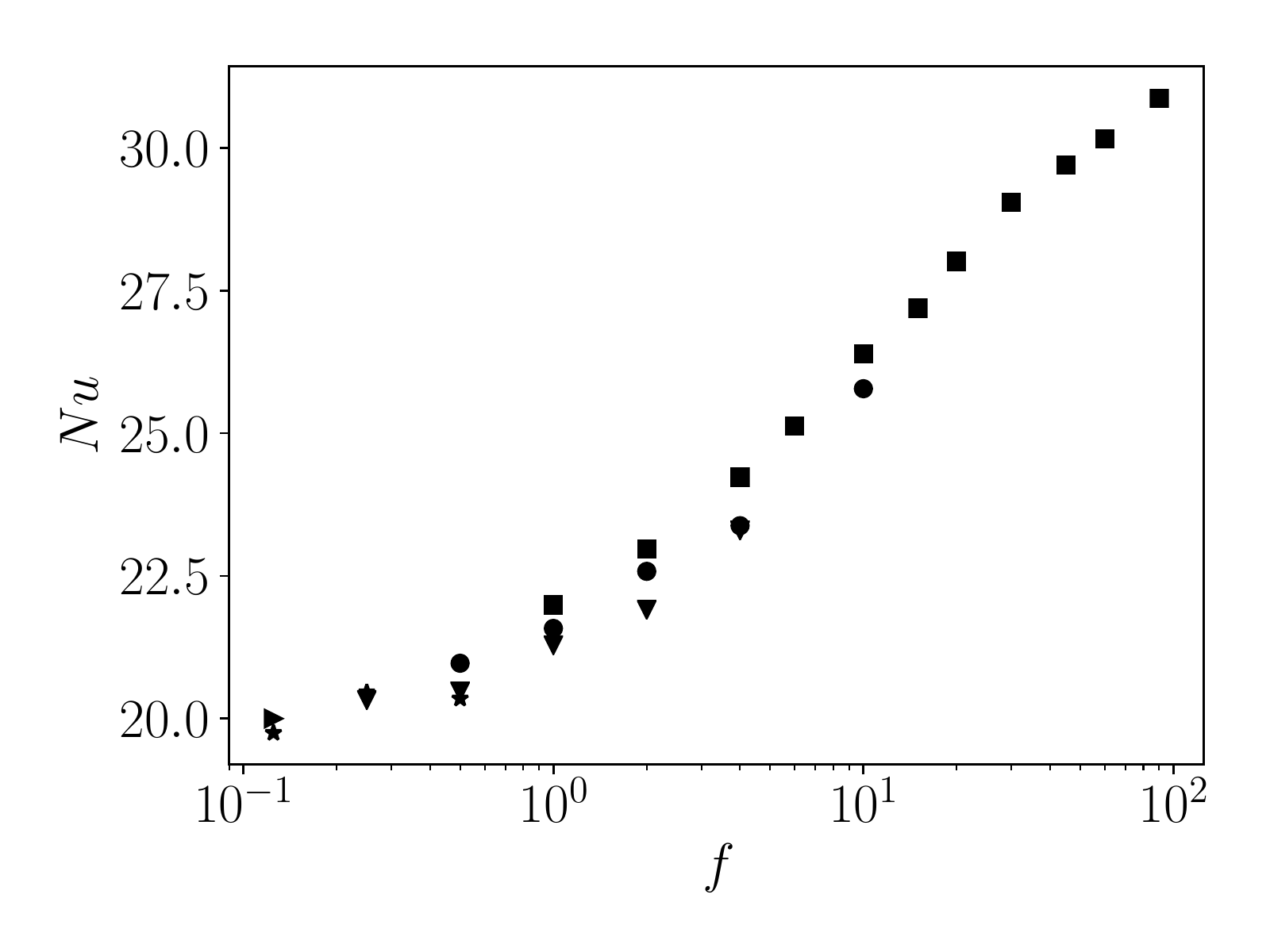}
 \includegraphics[width=0.48\linewidth]{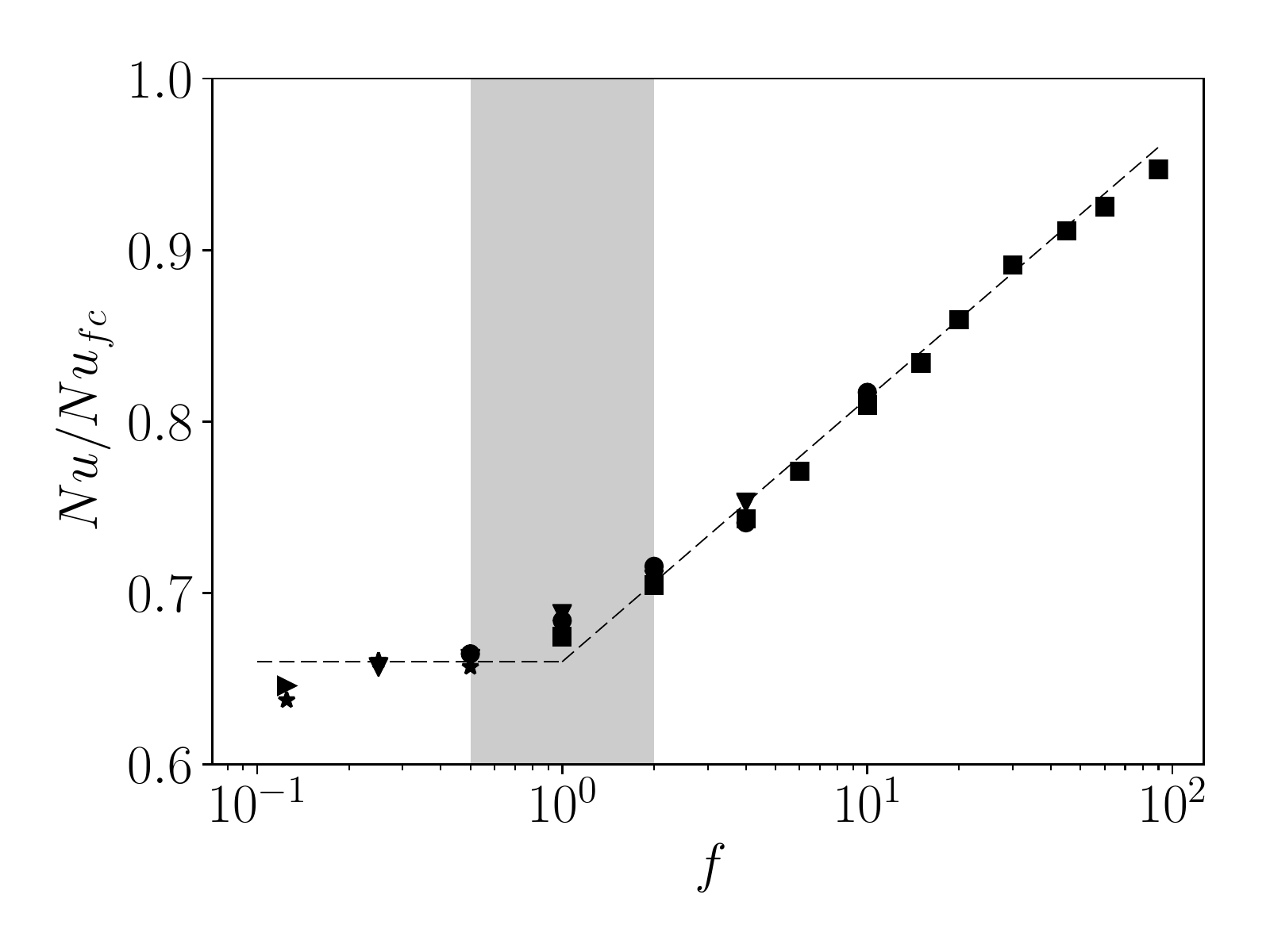}
  \includegraphics[width=0.48\linewidth]{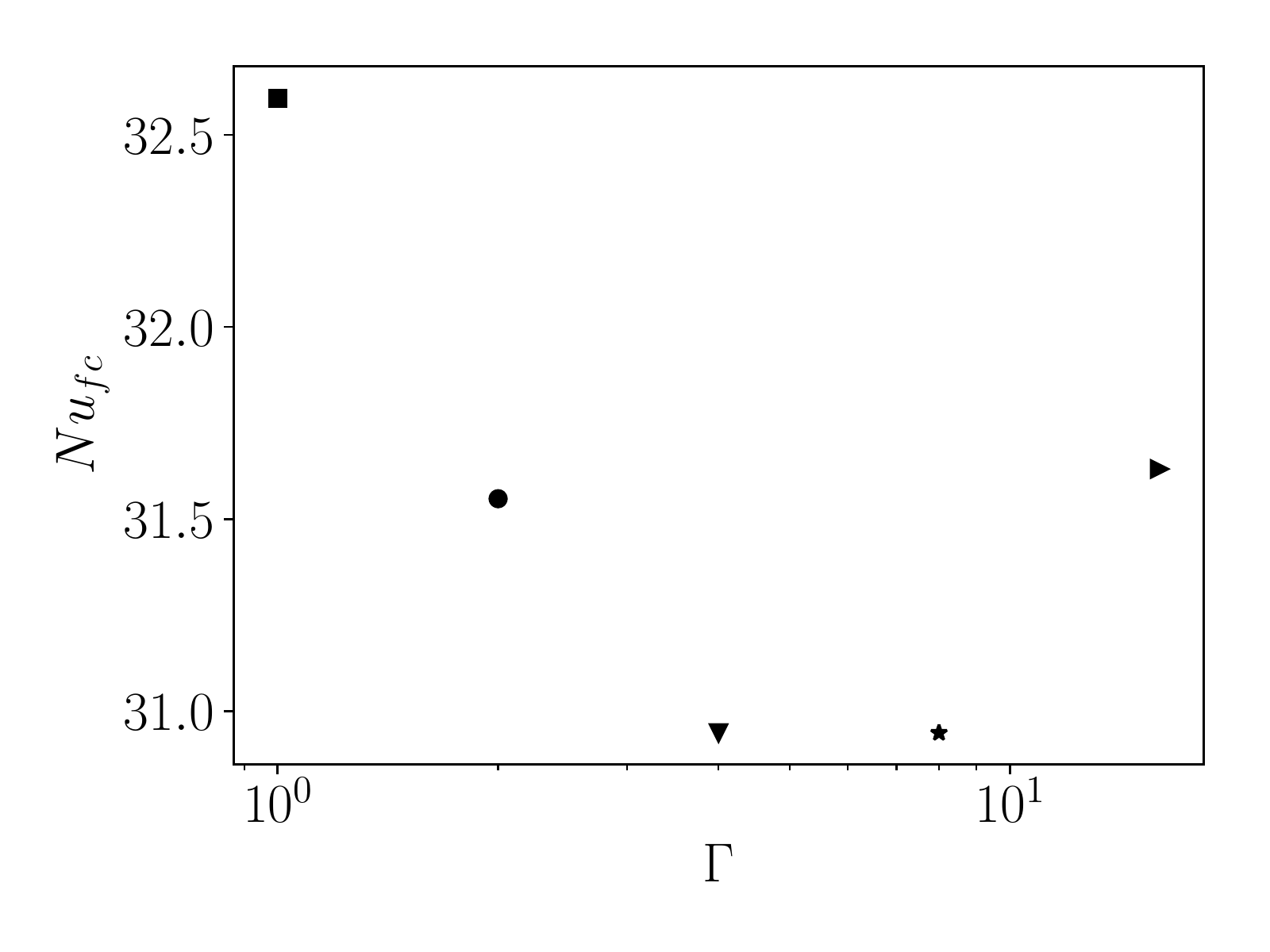}
 \includegraphics[width=0.48\linewidth]{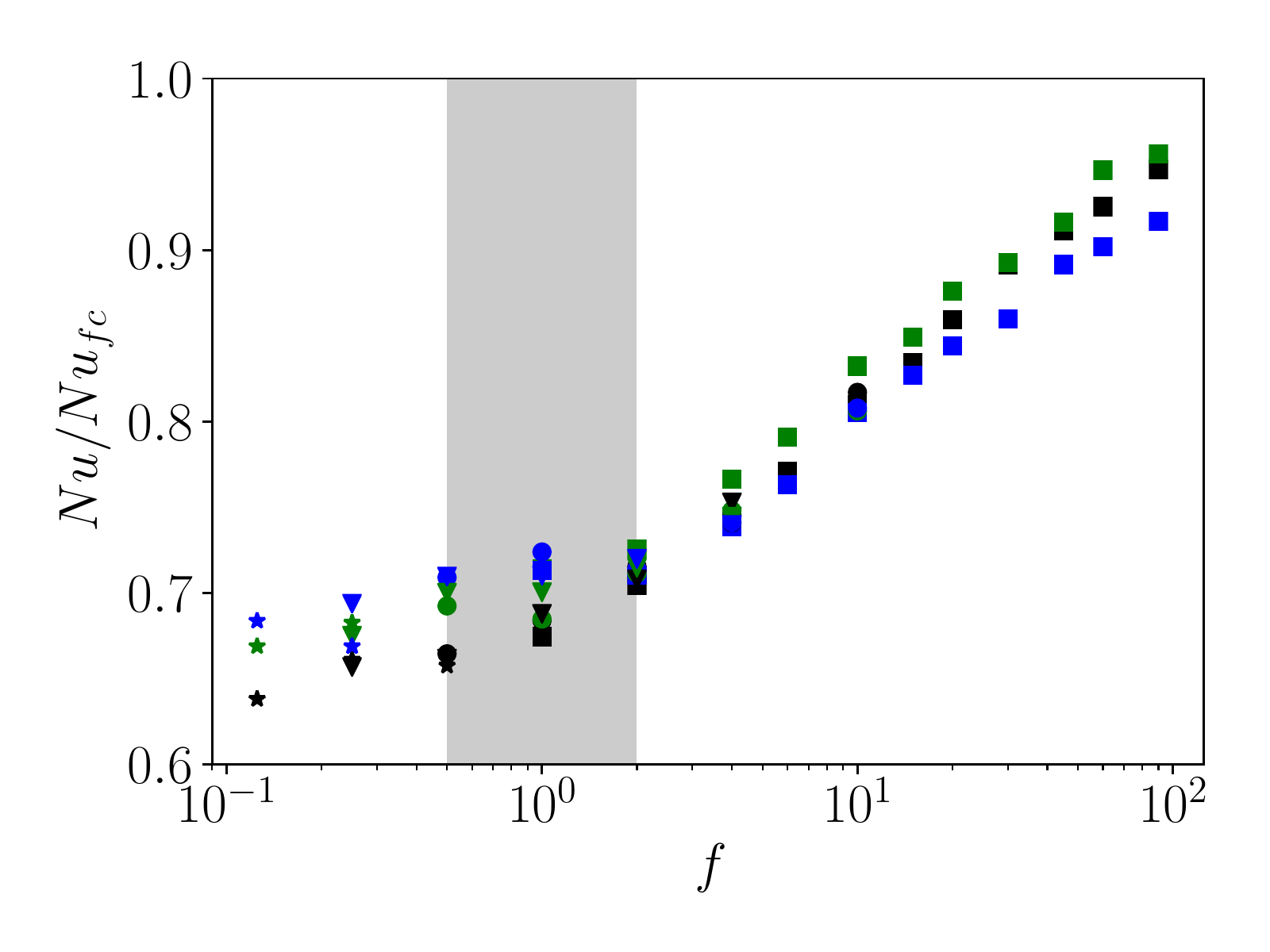}
 \caption{Top left: Nusselt number against stripe number for $Ra=10^8$ and $Pr=1$. A large variation between different values of $\Gamma$ can be seen. Bottom left: Nusselt number variation as a function of box-size for $Pr=1$. Top right: Normalized Nusselt number for $Pr=1$. The box-size dependence is removed when normalizing by a $\Gamma$-dependent fully conducting $Nu_{fc}$. Bottom right: Nusselt number normalized by the fully conducting value for each value of $\Gamma$. A large degree of collapse can be found across $Pr$. Symbols: color denotes Prandtl number, $Pr=1$ is black, $Pr=10$ is green and $Pr=100$ is blue. Symbol shape denote $\Gamma$: Squares are $\Gamma=1$, circles are $\Gamma=2$, V-triangles are $\Gamma=4$, stars are $\Gamma=8$ and $>$-triangles are $\Gamma=16$.}
 \label{fig:nu}
\end{figure}

We first focus on the heat transfer, non-dimensionalized as a Nusselt number $Nu$. The first step is to  eliminate box-size dependence $\Gamma$, such that $Nu(f)$ can be properly elucidated. As mentioned previously, we have conducted simulations for different aspect ratios $\Gamma$ with all other control parameters constant, to quantify as far as possible this dependence. The top-left panel of Figure \ref{fig:nu} shows the Nusselt number $Nu(\Gamma,f)$ against stripe frequency for $Pr=1$. The box-size dependence of $Nu$ can be appreciated in the fact that the data points for the same values of $f$ do not fall onto each other when $\Gamma$ is changed. This is not surprising, because from RB simulations with homogeneous boundary conditions, it is known that the Nusselt number shows some box-size dependency if $\Gamma \le 4$ \citep{ste18}. This dependence is further quantified in the bottom-left panel, where the Nusselt number of the fully-conducting (unmodified) system is shown as a function of $\Gamma$. While the variations are not very large, of the order of $6\%$, they are enough to introduce discrepancy into the measurements.

The $\Gamma$ dependence can be removed by adequately compensating $Nu$. In the top-right panel of Figure \ref{fig:nu}, we plot the normalized Nusselt number $Nu/Nu_{fc}(\Gamma)$, where $Nu_{fc}(\Gamma)$ is the Nusselt number for the homogeneous (fully conducting plates) case from the bottom-left panel. With this operation, we get an excellent collapse across all box-sizes, removing the $\Gamma$ dependence and elucidating the dependence on $f$. For values of $f<1$, the heat transport is almost constant, and independent of stripe wavelength. This is consistent with the observations by \cite{coo13,wan17}. For values of $f>1$, the heat transport begins to significantly increase, and tends towards the fully conducting value, consistent with \cite{bak18}.

For our simulations we are not able to reach $Nu/Nu_{fc}=1$ because we are limited in the values of $f$ we can simulate. This asymptotic value for $Nu/Nu_{fc}$ was reached in \cite{bak18} at lower Rayleigh numbers. For this to happen, the pattern size has to be much smaller than the thermal boundary layer size, implying notably finer grid resolution. If we estimate the thermal boundary layer size as $\lambda_T/H=1/(2Nu)$, this gives us the requirement of $f>>2Nu$ for $Nu/Nu_{fc}\to 1$. With our current Rayleigh number $Ra=10^8$, this should happen for $f>>65$. The largest value of $f$ simulated, $f=90$, is insufficient to see this.

In the top-right panel of Figure \ref{fig:nu}, we also shade in a postulated transition region $0.5<f<2$ between both regimes. No study of the aforementioned examined values on both sides of $f=1$, i.e.~they did not explore the transitional region between both behaviours, so they unsurprisingly reached different conclusions on the behaviour of the heat transport.

We attempt to extend this result for all Prandtl numbers simulated in the bottom-right panel of Figure \ref{fig:nu}, which shows the normalized Nusselt number for three different Prandtl numbers. While the basic result remains, i.e.~there is a sharp transition between two regimes of heat transport, the behaviour on both regimes appears slightly different. At first glance, the low-wavelength limit values of $Nu/Nu_{fc}$ are higher with increasing $Pr$, but it remains to be seen if these values would continue dropping as $f$ decreases, or if this is a product of the increased box-size dependency seen for larger Prandtl numbers \citep{ste18}. The slope of the $Nu/Nu_{fc}(f)$ relationship also appears to be steeper for higher $Pr$. The mechanisms for this are unclear, and any speculation on these trends is out of the scope of this study. What appears to be a general conclusion is that the wide-stripe and thin-stripe regimes exist independent of $Pr$ in our range of $1\le Pr \le 100$, and we expect them to persist as the geophysically-relevant regime $Pr\to\infty$ is entered.

\subsection{Temperature profiles}

The next thing we study is the dependence of the mean temperature at the mid-plane as a function of $f$. We do this by averaging the temperature in both horizontal directions, and in time (after the flow becomes statistically stationary). \cite{bak18} had concluded that as $f\to\infty$, the temperature in the mid-plane region would tend towards the arithmetic mean of the temperature of both plates, i.e.~$\theta=0.5$, and that as the stripes became larger, the mean temperature would tend towards two-thirds. This was rationalized by taking an average of the temperature at both plates, weighted by the fraction of the plate which actually conducts heat. For $\ell_C=0.5$, this gives $\theta=2/3$.

\begin{figure}
\centering
  \includegraphics[width=0.48\linewidth]{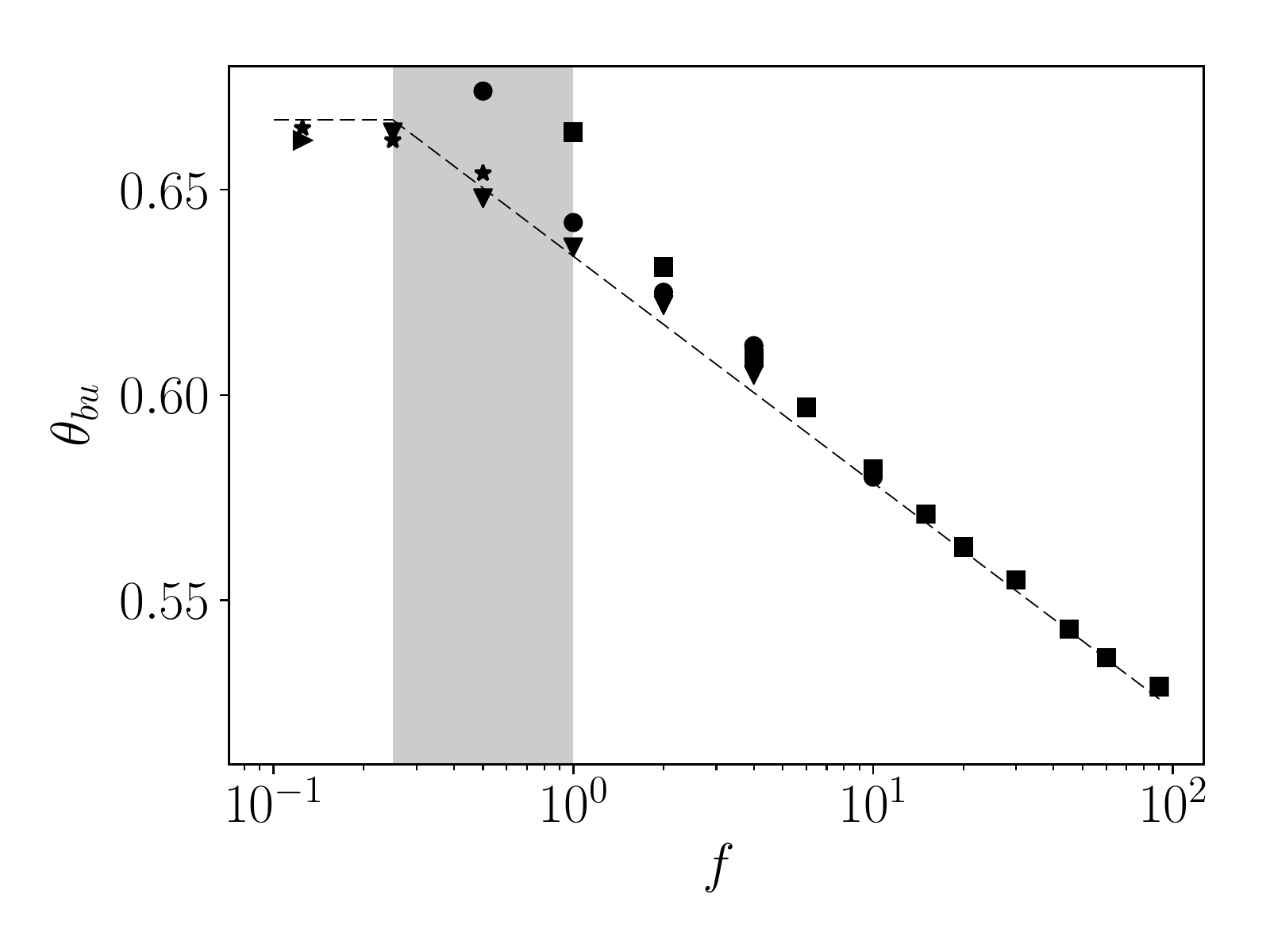}
  \includegraphics[width=0.48\linewidth]{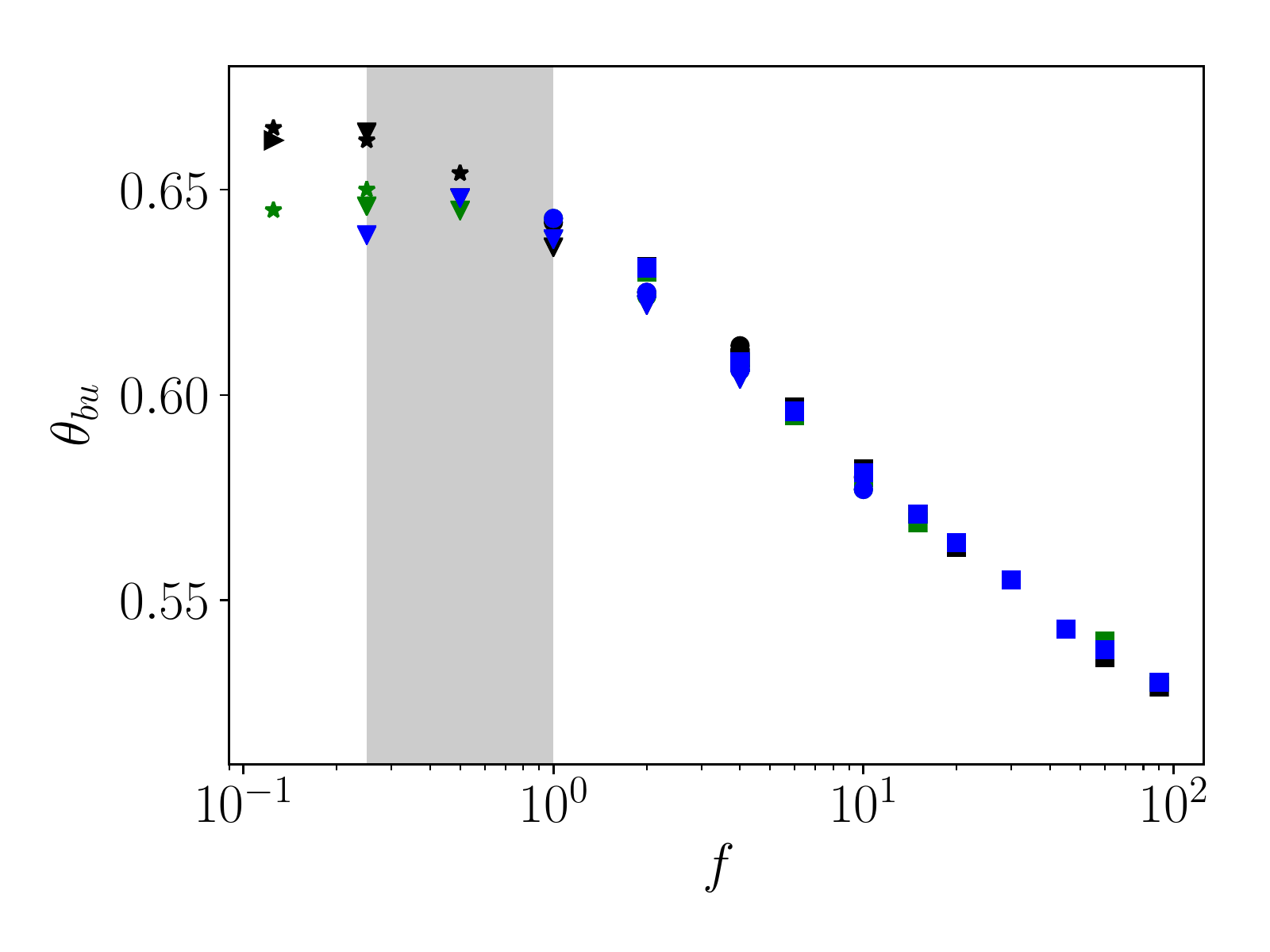}
  \caption{Left: Mid-gap temperature as a function of $f$ for $Pr=1$ and all box-sizes simulated. The box-size dependence of some simulations is emphasized. Right: Mid-gap temperature as a function of $f$ for all $Pr$ simulated and selected box-sizes. A clear transition between a constant-temperature region and a region of decreasing bulk-temperature can be seen. Shape and color of symbols are same as in Figure \ref{fig:nu}.}
  \label{fig:temp}
\end{figure}

The left panel of Figure \ref{fig:temp} shows the average mid-gap temperature as a function of $f$ for $Pr=1$. By analyzing the current simulations, it becomes clear that the simulations of \cite{bak18} have significant box-size dependence in their temperature measurements, due to the enforced periodicity constraining the flow and the development of large-scale structures. There is a strong box-size dependence in the transition region. When running a small periodic box with only one, or two stripes per simulation period, the temperature tends to increase due to the finite domain-size. Only when $\Gamma=4$ or $\Gamma=8$ are these effects mitigated, and we can obtain box-size independent results for $f\le 1$. As rationalized earlier, the mid-gap temperature can be seen to saturate to $\theta_{bu}=2/3$ for $f\le 0.25$, representing stripes of very large size. Consistent with \cite{bak18}, we see a sharp temperature drop as $f$ increases beyond $f=1$. We have shaded in a transitional region, which now happens for other values of $f$: $0.25<f<1$. To adequately capture this region, simulations of $\Gamma\ge 4$ are needed, which were not performed in \cite{bak18}.

In the right panel of Figure \ref{fig:temp} we show the bulk temperature as a function of $f$ for all the Prandtl numbers explored. Simulations which have a single unit-pattern per box-size are taken out of the results. By doing this, we see how the curves collapse in the high $f$ region. Again, there is a transition region which happens for $0.25<f<1$. The values of $f$ at which this transition happens are independent of $Pr$, but are different (smaller) than those from the transition we saw in $Nu/Nu_{fc}$.  Because of this, the low-$f$ region appears decreased in size in the figure.

\begin{figure}
\centering
  \includegraphics[width=0.48\linewidth]{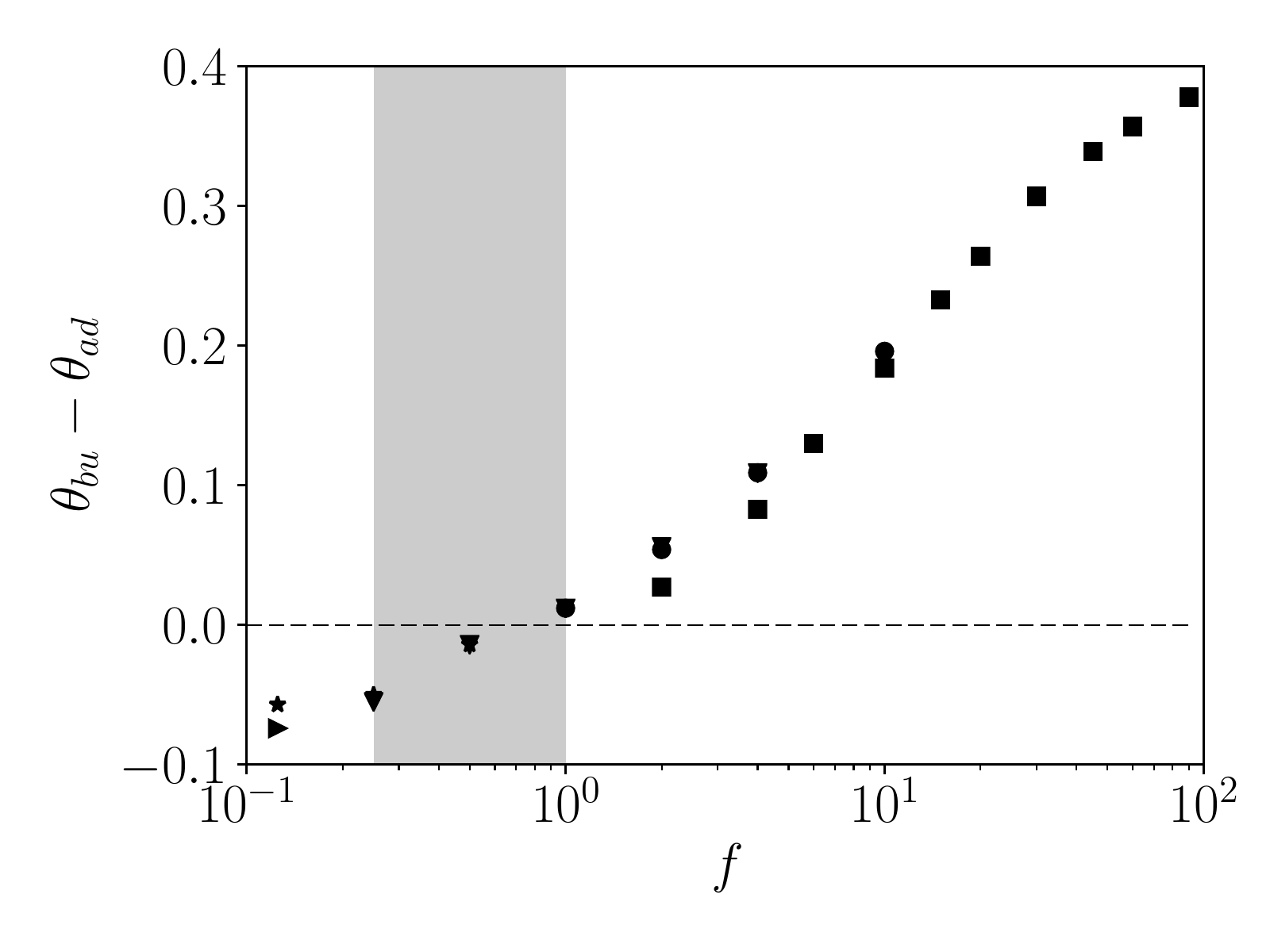}
  \includegraphics[width=0.48\linewidth]{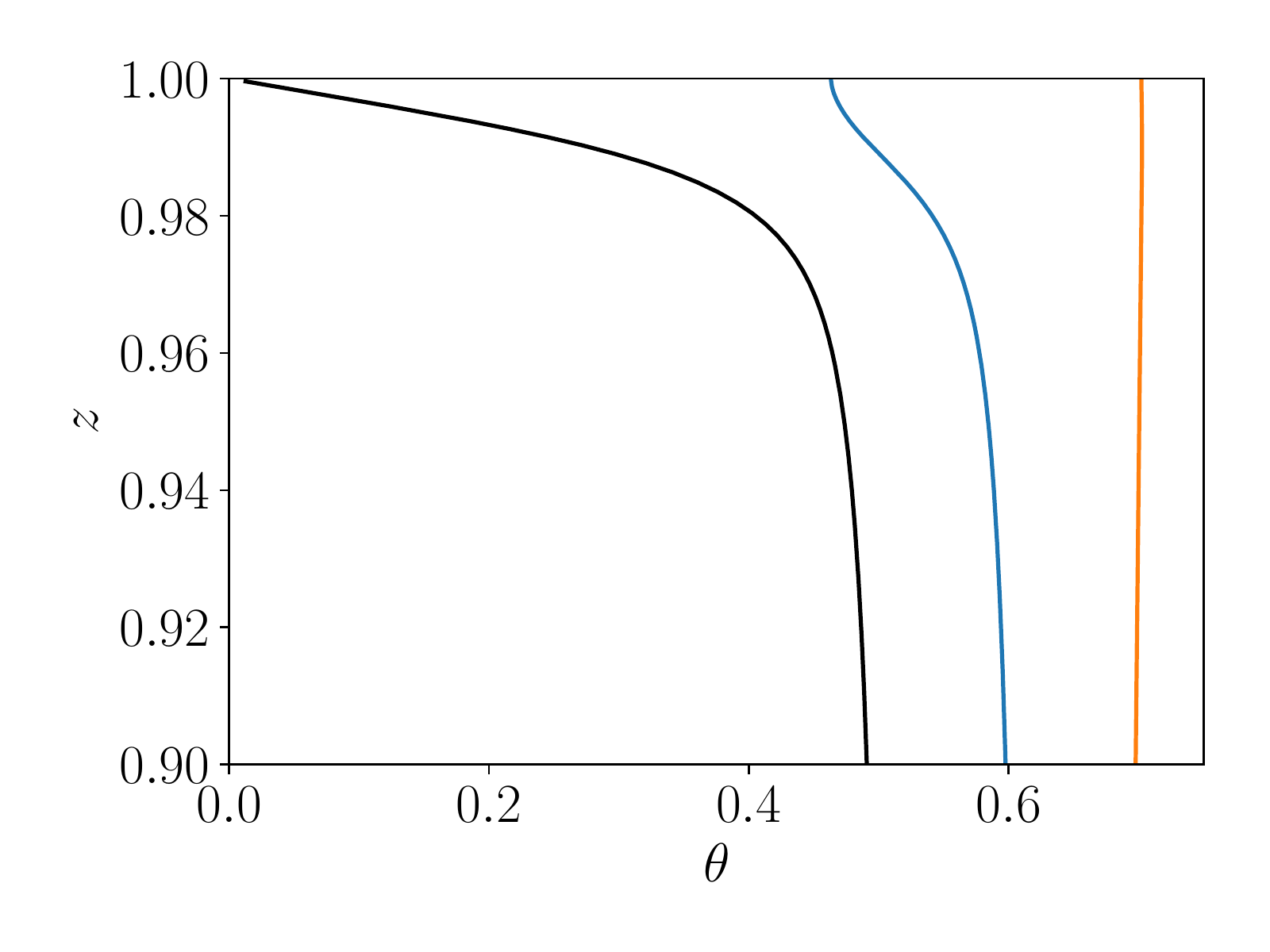}
  \caption{Left: Difference between the mid-gap temperature $\theta_{bu}$ and the average temperature of the fluid near the adiabatic plate as a function of $f$ for $Pr=1$ and all box-sizes simulated. Shape of symbols are same as in Figure \ref{fig:nu}. Right: thermal boundary layer profiles close to the adiabatic plates for $\Gamma=4$, $Pr=1$ and $f=4$ (blue), $f=0.25$ (orange), and the reference (homogeneous) case (black).}
  \label{fig:tempdiff}
\end{figure}

We can further explore the temperature statistics by analyzing the behaviour of the average fluid temperature close to the adiabatic plates ($\theta_{ad}$). Naively, we can expect the temperature of the adiabatic regions to tend towards the (cold) plate temperature $\theta=0$ as $f$ increases, and this is indeed what happens. What is more interesting is to look at the temperature difference between bulk and adiabatic region as a function of $f$, which we show in the left panel of Figure \ref{fig:tempdiff}. For large $f$, the bulk is hotter than the adiabatic regions at the top, cold plate, as expected. However, for $f<1$, the fluid close to the adiabatic regions becomes slightly hotter than the bulk fluid (as well as the conducting parts of the same plate), which means that there is a slight temperature inversion. This provides another statistic that shows a change in behaviour during the transition.

To further understand what is happening, we show the thermal boundary layer close to the adiabatic regions for a small $f$ and a large $f$ case in the right panel of Figure \ref{fig:tempdiff}. For the large $f$, we can see that inside the boundary layer, there is a temperature increase, which means that there is horizontal transport of heat to the conducting regions. For small $f$, this gradient is not seen, which means that the horizontal transport of heat within the boundary layers is severely weakened. This has the potential to generate strong horizontal flows due to strong density gradients, which we analyze in the next section.

\subsection{Velocity statistics and flow structure}

Finally, we analyze the velocity statistics and the changes in flow structure. As mentioned in the introduction, from the experimental and numerical results we expect that the large-stripe cases show a strong effect on the wind, while the small-stripe cases should not be very different from the canonical setup. We could also expect to see a strong asymmetry in the winds once the stripes pass a certain size, as excess heat must be redistributed horizontally from the zones close to adiabatic boundary conditions to the cooling parts.

To study the velocity statistics, we define the ``wind'' Reynolds number in the $i$-th direction as $Re_i=u^\prime_i H/\nu$, where $u^\prime$ is the volume averaged root-mean-square velocity. We also define the horizontal wind Reynolds number as $Re_h=\sqrt{Re_x^2+Re_y^2}$. For the homogeneous case, $u_x^\prime$ should be equal to $u_y^\prime$, meaning $Re_x=Re_y$. This means that $Re_h$ should be equal to $\sqrt{2}Re_x$, which is also equal to $\sqrt{2}Re_y$. However, due to the restrictions present due to the finite-domain (i.e.~finite $\Gamma$), $Re_x$ and $Re_y$ often differ from each other. Of all the statistics presented in this manuscript, $Re$ has the largest box-size dependence. This is true even for the case of fully conducting plates, as was shown by \cite{ste18}. For $Pr=1$, $Re_x$ and $Re_y$ differ from each other by about $5\%$, which gives us an estimate of the error made when measuring these quantities. For $Pr=10$ and $Pr=100$, the anisotropy is stronger, and causes deviations between $Re_x$ and $Re_y$ of over $100\%$. This means our error bars become too large to extract any meaningful information. Because of this, in this section we only show results for $Pr=1$.

\begin{figure}
  \includegraphics[width=0.48\linewidth]{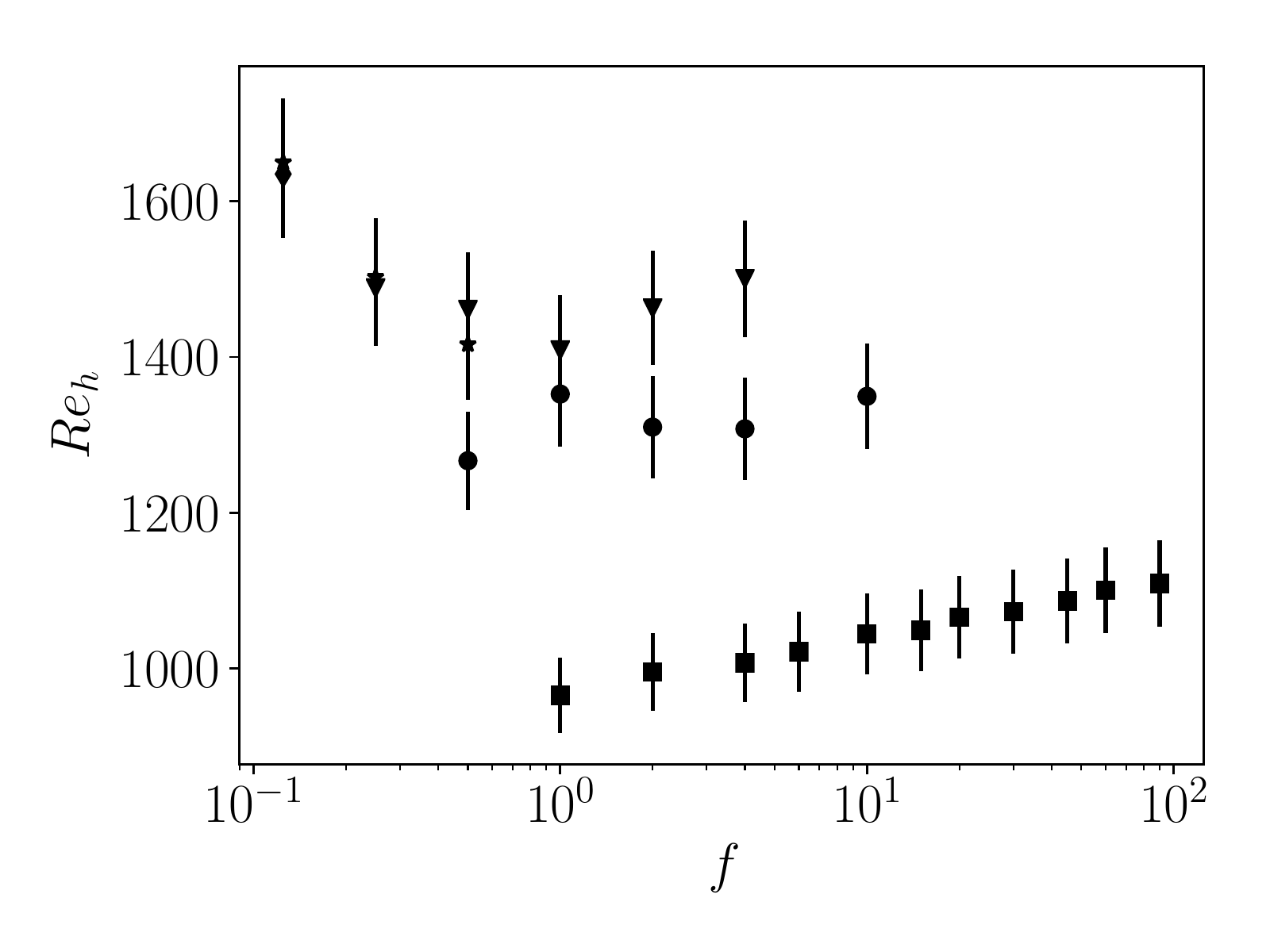}
  \includegraphics[width=0.48\linewidth]{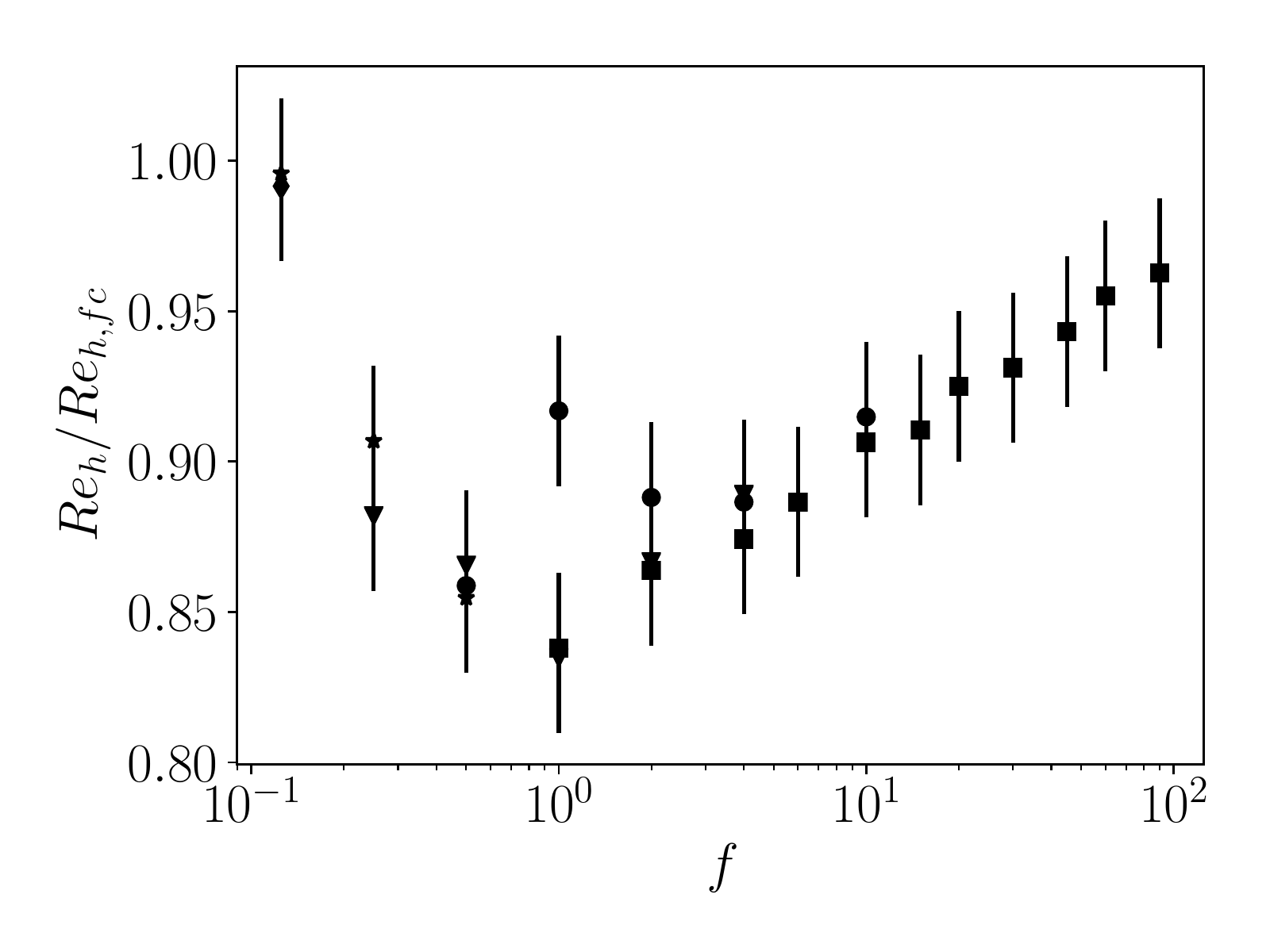}
  \caption{Left: Total horizontal flow strength quantified as a Reynolds number. Right; Total horizontal flow strength normalized by the homogeneous case. Symbol shape as in Figure 3. Error bars of $5\%$ are included on both plots.}
  \label{fig:reh}
\end{figure}

In the left panel of Figure \ref{fig:reh}, we show the horizontal Reynolds number $Re_h$ as a function of $f$ for the different values of $\Gamma$.  The dependence of $Re_h$ on $\Gamma$ is very strong: when comparing the values of $Re_h$ for the $f=10$ case at $\Gamma=1$ against the same $f$ at $\Gamma=2$, $Re_h$ can be seen to be as much as twice as large for $\Gamma=2$. By looking at both horizontal directions together, the underlying flow appears weaker than the homogeneous case. This is further quantified on the right panel, where we show $Re_h$ normalized by the fully conducting value $Re_{h,fc}$, which is $\Gamma$-dependent. From the figure we can observe a local minimum for $Re_h$ around $f=1$. As $f\to\infty$, we recover the homogeneous $Re_h$, and as $f$ becomes smaller than unity, the wind is enhanced by strong horizontal temperature gradients. For no cases simulated does the total wind strength exceed that of the homogeneous flow even if the data hints that this will happen for $f<0.1$. 
 
\begin{figure}
  \includegraphics[width=0.48\linewidth]{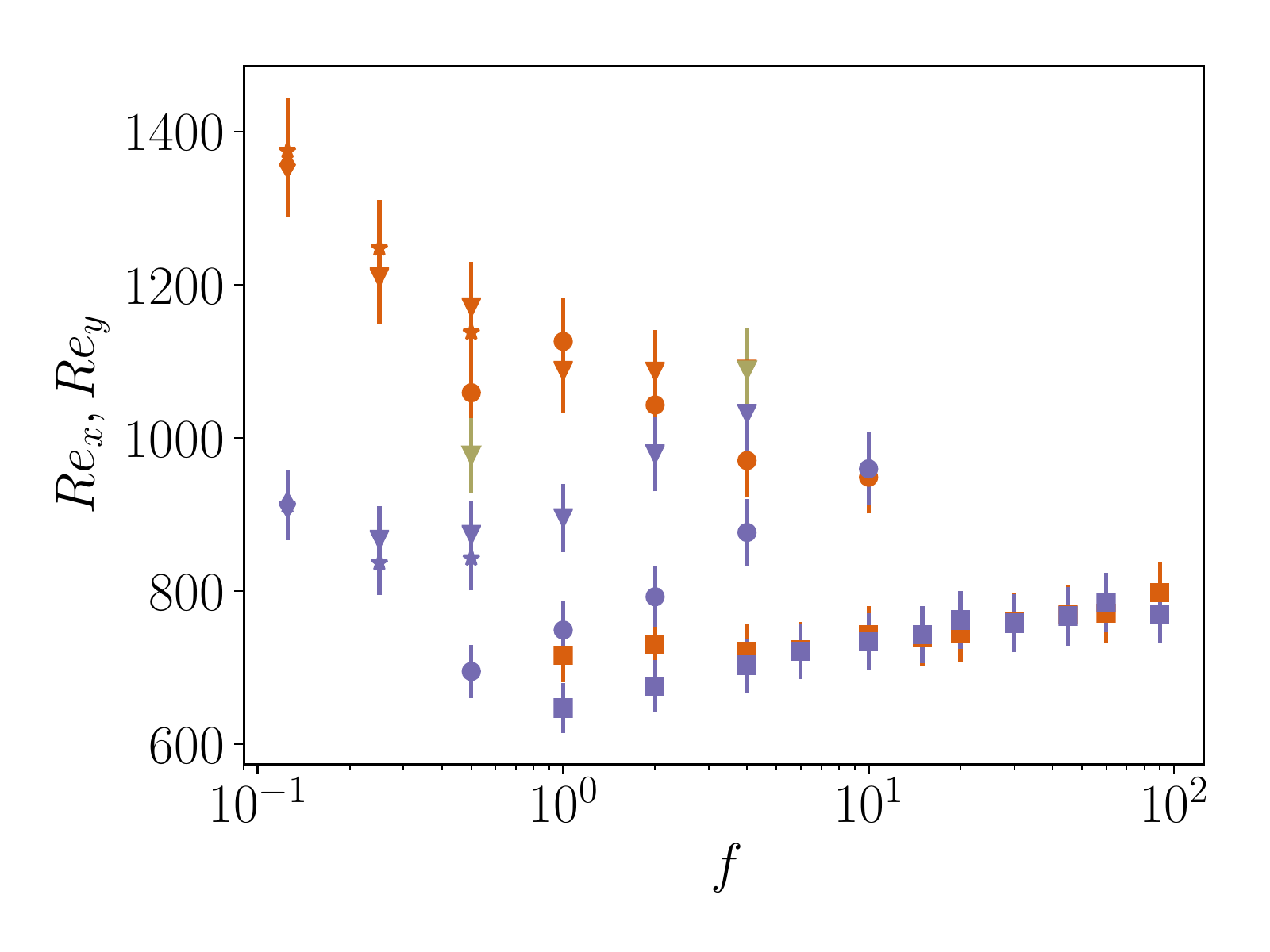}
  \includegraphics[width=0.48\linewidth]{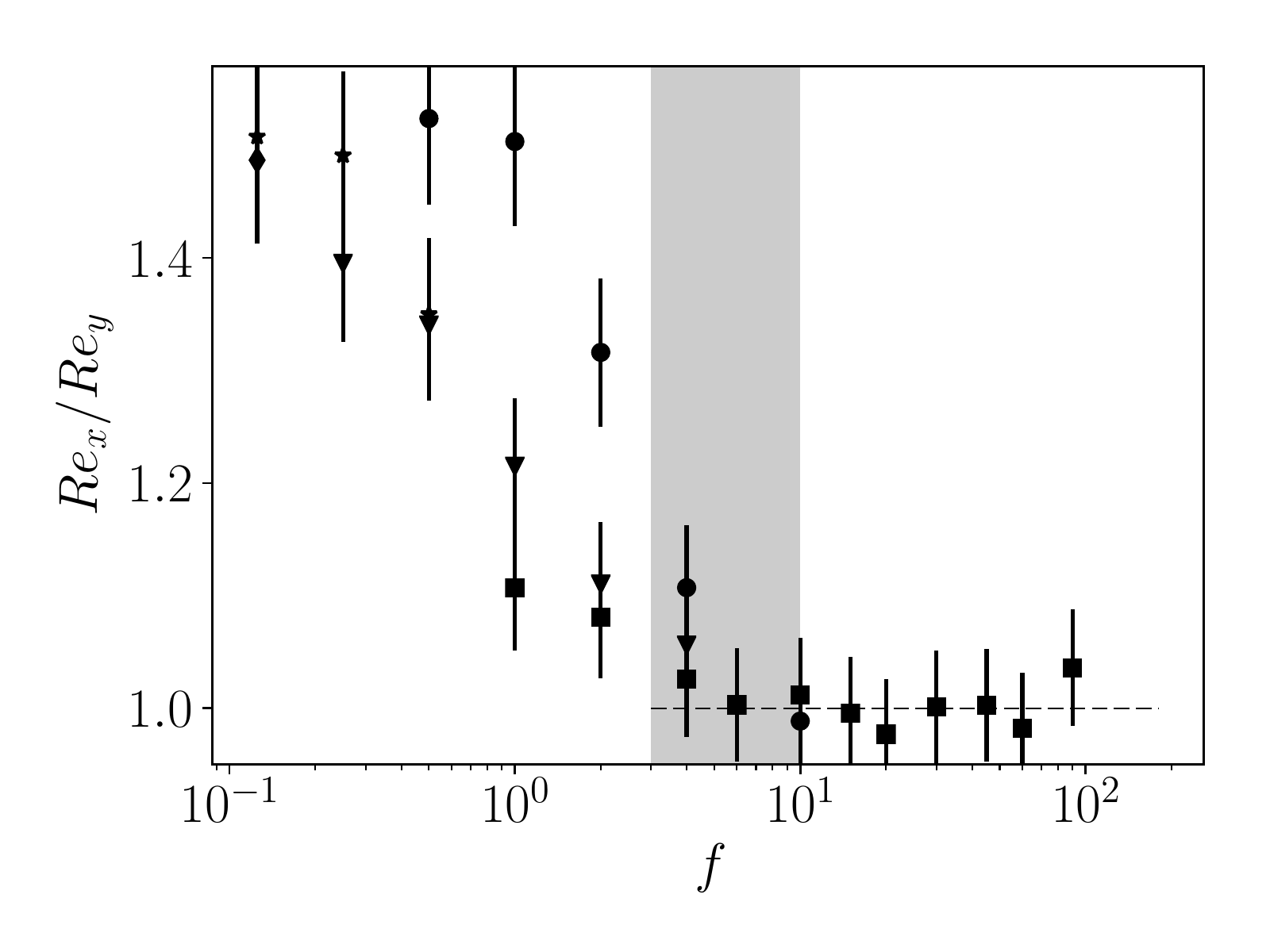}
  \caption{Left: Flow strength quantified as a Reynolds number in the stripe-normal ($Re_x$, orange symbols) and stripe-parallel ($Re_y$, purple symbols) directions. Data for the average horizontal Reynolds number ($\frac{1}{2}[Re_x+Re_y]$) for the checkerboard pattern simulations are included in khaki. Right: Ratio between the wind Reynolds numbers in the horizontal directions. Shape of symbols as in Figure \ref{fig:nu}. Error bars of $5\%$ are included on both plots. Only data for $Pr=1$ is presented here.}
  \label{fig:rms}
\end{figure}

We now turn to the behaviour of the individual $Re_i$ as a function of the stripe wavelength in figure \ref{fig:rms}, where we expect to see differences between both $x$ and $y$ directions due to the asymmetry introduced by the stripes. In the left panel of Figure \ref{fig:rms}, we show the large variability between the different values of $Re_x$ and $Re_y$ for different box-sizes and different values of $f$.  We can highlight a two trends. First, for large stripes, ($f<1$), the wind in the direction normal to the stripes, i.e.~ the $x$-direction is much larger than the wind in the direction parallel to the stripes, i.e.~the $y$ direction. As the stripes become smaller, the winds tend to equalize. In the direction parallel to the stripes, the wind increases, while in the direction perpendicular to the stripes, the wind decreases. This is further quantified in the right panel of Figure \ref{fig:rms}, which shows the ratio between $Re_x$ and $Re_y$. Once the stripes are small enough, around $f=4$, both Reynolds numbers become approximately equal, but they still are somewhat lower than the value for fully conducting plates, which is not reached until $f$ becomes much higher.

This asymmetric wind increase appears regardless of box size, even if there is some scatter on the magnitude of the asymmetry depending on $\Gamma$. Again, we have highlighted the transition region where the asymmetry disappears, which now occurs for higher values of $f$: $3<f<8$. We highlight that the transitional range for all of the three statistics happens for different values of $f$, even if they are generally around $f=1$.

We have also simulated two cases with a checkerboard pattern. This pattern restores the $x$-$y$ asymmetry, and serves to quantify how much of the wind enhancement comes from the necessity of transporting heat to the regions where it can be cooled by the plates instead of from the asymmetry of the system.  The checkerboard cases were ran for $\Gamma=4$ and $f=0.5$ and $f=4$. Unsurprisingly, we find that $Re_x\approx Re_y$ to within $5\%$, the previously reported error bound. What is more interesting is to compare the flow strength to the other case. This shows that the wind enhancement disappears with the flow asymmetry: for $f=0.5$ the value of the horizontal Reynolds number is lower than that for $f=16$, and this one is slightly below that of the homogeneous case. It is only when there is a strong asymmetry in the horizontal directions that intense large-scale flow patterns are generated in the direction perpendicular to the stripes, and the wind can exceed the value of the homogeneous case in a single direction.

As discussed in section 2 we only include results for $Pr=1$. For $Pr=10$ and $Pr=100$, a similar enhancement of the flow strength and anisotropy could be seen when applying the stripe pattern, but we cannot confidently attribute it to physical reasons alone. For this reason we decided not to show this data. Further studies at increasing box-size at high $Pr$ are required to adequately disentangle the box-size effects from those coming from inhomogeneous boundary conditions.

\section{Conclusions}

We have conducted numerical simulations of adiabatic-conducting stripe pairs of Rayleigh-B\'enard flow in an attempt to uncover the differences between the large-stripe regime of \cite{coo13, wan17}, and the small-stripe regime of \cite{rip14,bak18}. We observed that for $Ra=10^8$ and $Pr>1$, a transition between the two behaviours happens at around $f=1$, that is, stripes the size of the plate distance. We note that the statistics show transitions at different values of $f$, and that these transitions can only be fully uncovered once the simulation ``box-size'' is large enough, but as a rule of thumb, they happen around $f\approx 1$. While this is not totally satisfactory, we point out that for rotating Rayleigh-B\'enard, we have previously seen different statistics such as fluctuations, or mean temperature profiles, have different transition points \citep{kun16}.

The small-stripe regime is characterized by a heat transport that is very dependent on stripe-wavelength. With decreasing stripe wavelength, the bulk temperature and heat transport asymptotically tend to the fully conducting values, even when only half the plate is conducting. This regime was already explored in detail by \cite{bak18}. Only when the stripes become very small, i.e.~$f>>2Nu$, do the statistics converge to those of fully conducting plates.

The large-stripe regime is characterized by a heat transport and a bulk temperature that is independent of stripe wavelength. To maintain the heat transport, even as the stripes become large, the underlying flow is heavily modified (as observed by \cite{wan17}), and the velocities in the stripe-normal direction increase substantially with increasing stripe-length. The temperature stabilizes at values around $2/3$, i.e. the weighted average of the plate temperatures, but this appears to depend on Prandtl number. There appears to be a local temperature inversion, where the temperature at the adiabatic plates is higher than the bulk temperature. This regime is geophysically interesting, as many variations on the Rayleigh-B\'enard problem such as mantle convection consist of very long wavelength inhomogeneities. We confirm here that these types of variations can lead to the enhancement of underlying ``winds'' or circulations.

However, the conclusions here are limited in two main ways. The simulation of large $Pr$ flows is complicated due to the resolution requirements, and the computational box-sizes. The effect of $Pr$ on enhancing winds has to be quantified in a more detailed manner. In second place, the effect of $l_C$, i.e.~the ratio of conducting to adiabatic areas has been fixed at one half. Repeating this study for other values of $l_C$ is necessary to achieve more robust conclusions.

{\em Acknowledgments:} We thank A. Blass, D. Bakhuis and R. J. A. M. Stevens for providing their data. We thank the HPE Data Science Institute (HPE DSI) at the University of Houston for providing computing resources. 

\bibliographystyle{jfm}
% Note the spaces between the initials
\bibliography{main}

\end{document}